# Scalable coarse-to-fine spatial downscaling

Daisuke Murakami[1], Yongwan Chun[2], Takahiro Yoshida[3], Hajime Seya[4]

[1] Department of Fundamental Statistical Mathematics, Institute of Statistical Mathematics, Tachikawa, Tokyo, Japan

(E-mail: dmuraka@ism.ac.jp)

[2] School of Economic, Political and Policy Sciences, The University of Texas at Dallas, Richardson, Texas, USA

[3] Center for Spatial Information Science, The University of Tokyo, Japan

[4] Graduate School of Engineering Department of Civil Engineering, Kobe University, Japan

Abstract: This study proposes coarse-to-fine downscaling (CF-DS), a scalable spatial downscaling method extending coarse-to-fine spatial modeling. Unlike conventional spatial-statistical downscaling methods such as area-to-point kriging, CF-DS does not require covariance matrix inversion or likelihood evaluation. Instead, it represents latent spatial processes through the synthesis of multi-scale local models, substantially reducing computational cost while approximately satisfying the aggregation constraint. Monte Carlo experiments show that CF-DS achieves predictive accuracy comparable to area-to-point kriging with dramatically shorter computation times, particularly for large datasets. An application to downscaling electricity consumption in the Tokyo metropolitan area further demonstrates its practical usefulness. The results suggest that CF-DS provides an efficient alternative for large-scale spatial downscaling problems. CF-DS is implemented in an R package spCF (https://cran.r-project.org/web/packages/spCF/index.html).

# 1. Introduction

The spatial support of geographic data (i.e., aggregation units or observation locations) is diverse, and its typical form differs across disciplines. In regional science, data are typically aggregated into administrative units, whereas data in the natural sciences are typically point-referenced or aggregated into regular grids. In practice, analysis often requires combining datasets across disciplines (see Gelfand and Schliep, 2026), but it is not straightforward to simultaneously handle data observed at different aggregation units or spatial locations. In addition, available data are often aggregated at spatial units that are too coarse to meet the resolution requirement for the analysis. In response to these challenges, the change of support problem (COSP), which converts the spatial support of data into another support that better meets the purpose of an analysis, has long been studied (see Gotway and Young, 2002).

Downscaling (e.g., Atkinson, 2013, Wang et al., 2015), which redistributes data aggregated at a coarse spatial unit to finer units, has been studied as a remedy for the COSP. Downscaling not only increases the resolution of the data but also enables re-aggregation to the spatial units better suited to the analysis purpose. While a wide variety of downscale methods have been developed to accurately predict fine-scale patterns from aggregated data, some recent approaches rely on disaggregated training data to learn relationships between aggregated and disaggregated patterns (e.g., Vandal et al., 2017; Stengel et al., 2020). However, such disaggregated data are often unavailable or difficult to obtain in practice. Therefore, this study focuses on downscaling methods that infer disaggregate-level patterns solely from observed aggregated values and disaggregated auxiliary variables, without requiring disaggregated values.

Representative downscaling methods developed for such settings can be classified into (i) deterministic methods, (ii) spatial-process-based methods, and (iii) machine-learning methods. Specific examples include: (i) areal weighting interpolation (e.g., Flowerdew and Green, 1994), dasymetric mapping (Fisher and Langford, 1996), and the point-in-polygon method (e.g., Sadahiro, 2000); (ii) area-to-point (ATP) kriging (Kyriakidis, 2004; Wang et al., 2015) and hierarchical Bayesian models (Mugglin et al., 1999); (iii) random forests (e.g., Stevens et al., 2015; Chen et al., 2024) and neural networks (e.g., Ulyanov et al., 2018). Most of these methods have been developed to satisfy an aggregation constraint that the aggregated values of the disaggregate-level predictions match the observed aggregated values.

The accuracy of each approach depends on their settings. (i) Deterministic methods tend to be effective when auxiliary variables that strongly explain the target data are available. For example, when building-level residential total floor-area data are available, fine-scale population distributions can be predicted accurately by dasymetric mapping using total floor areas as the distribution weights (e.g., Wang et al. 2021). Spatial process-based methods (ii) explicitly model the disaggregate-level processes. It is well suited to downscaling spatially smooth variables such as temperature. By contrast, machine-learning methods (iii) predict disaggregate-level patterns by learning potentially non-linear relationships between aggregated values and disaggregated auxiliary variables.

Among these, spatial process-based methods (ii) have several advantages. It yields stable predictions even when auxiliary data are limited, predicted values do not change discontinuously at the boundaries of aggregation units, and the predictive uncertainty can be quantified. However, these methods tend to be computationally intensive (Wang et al., 2015), and their applications have been limited to relatively small-scale problems. Although computationally efficient approximations have been proposed (e.g., Forlani et al., 2020; Moraga and Alahmadi, 2026; Rodriguez Avellaneda et al., 2026), a considerable room remains for improving both downscaling accuracy and computational efficiency.

The objective of this study is to develop a downscaling method that explicitly models a disaggregate-level spatial process with much smaller computation cost than conventional methods. To this end, we extend the coarse-to-fine spatial modeling (CFSM; Murakami et al., 2026a,b) framework and propose a new downscaling method, termed coarse-to-fine downscaling (CF-DS). Unlike conventional spatial process models, which are represented through covariance modeling, CFSM represents spatial processes by synthesizing local models across a sequence of spatial scales, and CF-DS inherits this framework. As shown later in this paper, this property enables highly scalable downscaling while preserving the ability to capture disaggregated spatial patterns.

The remainder of this paper is organized as follows. Section 2 describes the problem setting, including the aggregation constraint. Section 3 develops the proposed CF-DS method, and Section 4 evaluates its predictive accuracy and computational efficiency through Monte Carlo experiments. Section 5 applies it to downscale commercial electricity consumption in the Tokyo metropolitan area, Japan. Lastly, Section 6 concludes our discussion.

# 2. Problem setting

We consider predicting the unobserved responses $y_1, \ldots, y_n$, at the disaggregated units (index: $i \in \{1, \ldots, n\}$) from the observations $Y_1, \ldots, Y_N$ at the aggregated units (index: $I \in \{1, \ldots, N\}$; $n > N$).

The response variable is assumed to be either extensive data, such as counts including population or non-integer additive quantities such as electricity consumption, or intensive data, such as population density, average income, and other spatially averaged variables.

For extensive data, $Y_I$ and $y_i$ must satisfy the following aggregation constraint:

$$Y_I = \sum_{i \in I} y_i. \tag{1}$$

where $i \in I$ denotes the disaggregated units contained in the aggregated unit *I*. Eq. (1) simply states that the sum of the values within each aggregated unit equals to its observed value. For intensive data $Y_I^{(intens)}$, the aggregation constraint takes the following form:

$$Y_I^{(intens)} = \sum_{i \in I} \frac{a_i}{A_I} y_i^{(intens)}. \tag{2}$$

where $A_I = \sum_{i \in I} a_i$. Here, $a_i$ is the quantity corresponding to the denominator of the intensity at disaggregate unit *i*. For example, if $y_i^{(intens)}$ is the number of crimes per unit area, then $a_i$ is area; if it is the number of crimes per capita, then $a_i$ is population. Eq. (2) means that the weighted average of the disaggregated values equals to the aggregated value[1]. In what follows, we proceed assuming extensive data, but the intensive-data case can be formulated analogously; see Appendix 1 for details.

In downscaling of extensive data, proportional allocation according to a proportional weight $a_i$ has often been adopted, as follows:

[1] $a_i$ is the deviser to convert extensive data into intensity. Intensive data are expressed as $y_i^{(intens)} = y_i / a_i$ at the disaggregate level and $Y_I^{(intens)} = \frac{\sum_{i \in I} y_i}{\sum_{i \in I} a_i}$ at the aggregate level. Eq. (2) must therefore hold for them, and substitution gives $\sum_{i \in I} \frac{a_i}{\sum_{i \in I} a_i} y_i^{(intens)} = \frac{\sum_{i \in I} y_i}{\sum_{i \in I} a_i} = Y_I^{(intens)}$ confirming that Eq. (2) holds.

$$\hat{y}_i = \frac{a_i}{A_I} Y_I. \tag{3}$$

This proportional allocation is called dasymetric mapping (Fisher and Langford, 1996) in geography and is also commonly called as areal weighting interpolation when $a_i$ represents area. Eq. (1) is satisfied regardless of the choice of $a_i$; it is confirmed by aggregating $\hat{y}_i$ as $\sum_{i \in I} \hat{y}_i = \sum_{i \in I} \frac{a_i}{\sum_{i \in I} a_i} = Y_I$. In spatial process-based methods such as ATP kriging, best linear unbiased predictors satisfying Eq. (1) have been used. However, these methods require the processing of an $n \times n$ covariance matrix and computation of $N \times N$ inverse matrices, which makes them unsuitable for large-scale problems (Wang et al., 2015). Given the growing demand for large-scale applications, such as population downscaling to 1-km grids covering the entire globe in climate science and related fields (e.g., Stevens et al., 2015; Tomari et al., 2026), this property is a major practical obstacle.

# 3. Proposed method

To address the computational challenges of conventional spatial process-based methods, this study extends CFSM to develop a novel downscaling method called CF-DS as a scalable alternative that is better suited for large-scale downscaling applications. The remainder of this section is organized as follows. Section 3.1 introduces the model specification. Section 3.2 presents the local parameter estimation. Section 3.3 derives the predictor and discusses its aggregation consistency. Section 3.4 describes the learning algorithm.

## 3.1 Model

### 3.1.1 Data model

The following disaggregated-level model is assumed for the unknown response variable $y_i$:

$$y_i = \sum_{k=1}^{K} a_i x_{i,k} \beta_k + z_{i,1:R} + e_i, \quad e_i \sim N\left(0, \sigma^2 a_i^2\right), \tag{4}$$

where $x_{i,k}$ is a covariate, $\beta_k$ is a coefficient, $e_i$ represents noise, and $\sigma^2$ is a variance parameter. $a_i$ corresponds to the proportional weight used in dasymetric mapping and is given by a variable assumed to be proportional to the response. $z_{i,1:R}$ denotes the spatial process, which is described in Section 3.1.2. The aggregated-level model is given by aggregating each term in Eq. (4) using Eq. (1) as follows:

$$Y_I = \sum_{i \in I} \sum_{k=1}^{K} a_i x_{i,k} \beta_k + Z_{I,1:R} + E_I, \qquad E_I \sim N\left(0, \sigma^2 \sum_{i \in I} a_i^2\right), \tag{5}$$

where $Z_{i,1:R} = \sum_{i \in I} z_{i,1:R}$.

### 3.1.2 Process model

In CFSM, $z_{i,1:R}$ is represented as the following multi-scale process:

$$z_{i,1:R} = \sum_{r=1}^{R} b_r z_{i,r}, \tag{6}$$

where $b_r$ represents coefficients. $z_{i,1}, \ldots, z_{i,R}$ are scale-wise processes corresponding to bandwidths $h_1, \ldots, h_R$ satisfying $h_r = \alpha h_{r-1,}$ with $0 < \alpha < 1$. In other words, $z_{i,r}$ represents the *r*-th largest-scale process. The number of scales $R$ is treated as unknown.

The $r$-th single-scale process $z_{i,r}$ is constructed by synthesizing $C_r$ local models distributed across the study area. The $c_r$-th local model, centered at $c_r$-th local center, is defined as follows:

$$z_{i,r}|c_r \sim N\left(a_i\mu_{c_r}, \frac{a_i^2 v_{c_r}^2}{w_{h_r}^2(d_{i,c_r})}\right). \tag{7}$$

where $z_{i,r}|c_r$ represents $z_{i,r}$ conditioned on the model assuming the local mean $a_i\mu_{c_r}$ and variance $a_i^2 v_{c_r}^2$ at the center. Following dasymetric mapping, the mean is assumed to vary in proportion to the proportional weight $a_i$. $w_{h_r}(d_{i,c_r})$ represents the local kernel that decays in accordance with the distance $d_{i,c_r}$ between site $i$ and the $c_r$-th local center. This study assumes $w_{h_r}(d_{i,c_r}) = \exp(-d_{i,c_r}/h_r)$ following Murakami et al. (2026a).

The predictive mean and variance of $z_{i,r}|c_r$ are

$$E[z_{i,r}|c_r] = a_i\mu_{c_r}, \qquad V[z_{i,r}|c_r] = \frac{a_i^2 v_{c_r}^2}{\sum_{i=1}^{n} w_{h_r}^2(d_{i,c_r})} + \frac{a_i^2 v_{c_r}^2}{w_{h_r}^2(d_{i,c_r})}. \tag{8}$$

The spatial process $z_{i,r}$ is constructed by synthesizing the $C_r$ local models. It is achieved by taking the product of their probability density functions.[2] The resulting $z_{i,r}$ yields the following single-scale process:

$$z_{i,r} \sim N(\hat{z}_{i,r}, \hat{v}_{i,r}^2), \tag{9}$$

[2] This synthesis minimizes the average Kullback–Leibler divergence from the local models to the synthesized distribution, so that the tendencies captured by the individual local models are reflected as much as possible (Cao and Fleet, 2015).

$$\hat{z}_{i,r} = \hat{v}_{i,r}^2 \sum_{c_r=1}^{C_r} \frac{E[z_{i,r}|c_r]}{V[z_{i,r}|c_r]}, \quad \hat{v}_{i,r}^2 = 1 \Big/ \sum_{c_r=1}^{C_r} \frac{1}{V[z_{i,r}|c_r]}.$$

This formulation is identical to the original CFSM except for the introduction of the proportional weight $a_i$. The next section explains the estimation of $\mu_{c_r}$ and $v_{c_r}^2$, required to obtain $z_{i,r} \sim N(\hat{z}_{i,r}, \hat{v}_{i,r}^2)$.

## 3.2 Local parameter estimates

The local model (Eq. 7) is aggregated using Eq. (1) as follows:

$$Z_{I,r}|c_r \sim N(A_I \mu_{c_r}, v_{c_r}^2 V_{I,c_r}), \tag{10}$$

where $Z_{I,r} = \sum_{i \in I} z_{i,r}$ and $V_{I,c_r} = \sum_{i \in I} \frac{a_i^2}{w_{h_r}^2(d_{i,c_r})}$.[3] If $Z_{I,r}$ is known, the local parameters $\mu_{c_r}$ and $v_{c_r}^2$ can be estimated by the weighted least squares method as follows:

$$\hat{\mu}_{c_r} = \frac{\sum_I A_I V_{I,c_r}^{-1} Z_{I,r}}{\sum_I A_I^2 V_{I,c_r}^{-1}}, \qquad \hat{v}_{c_r}^2 = \frac{1}{(N_{c_r} - 1)} \sum_{I=1}^{N} \frac{\left(Z_{I,r} - A_I \hat{\mu}_{c_r}\right)^2}{V_{I,c_r}}, \tag{11}$$

where $N_{c_r}$ is the number of sample sites satisfying $w_{h_r}(d_{i,c_r}) > 0$. The predictive mean and variance of $z_{i,r} \sim N(\hat{z}_{i,r}, \hat{v}_{i,r}^2)$ can be evaluated by substituting $\hat{\mu}_{c_r}$ and $\hat{v}_{c_r}^2$ into Eqs. (8) and (9).

---

[3]Because $\mu_{c_r}$ does not depend on $i$, $\sum_{i \in I} a_i \mu_{c_r} = \mu_{c_r} \sum_{i \in I} a_i = A_I \mu_{c_r}$. Furthermore, since $a_i$ is given, $V[E_{I,c_r}] = \sum_{i \in I} a^2(s_i) V[e_{c_r}(s_i)] = \sum_{i \in I} \frac{a^2(s_i)\sigma_{c_r}^2}{w_{h_r}^2(d_{i,c_r})} = \sigma_{c_r}^2 V_{I,c_r}$.

## 3.3 Predictor and aggregation constraint

The predictor of $y_i$ is obtained by taking its expectation given Eqs. (4), (6), and (9), as follows:

$$\hat{y}_i = \sum_{k=1}^{K} a_i x_{i,k} \beta_k + \hat{z}_{i,1:R}, \qquad \hat{z}_{i,1:R} = \sum_{r=1}^{R} b_r \hat{z}_{i,r}, \tag{12}$$

Below, we examine the conditions under which $\hat{y}_i$ satisfies the aggregation constraint. We first note that $\hat{z}_{i,r}$ (Eq. 9) can be rewritten as follows:

$$\hat{z}_{i,r} = \sum_{c_r=1}^{C_r} \widetilde{W}_{i,c_r} \sum_{I=1}^{N} \widetilde{w}_{I,c_r} Z_{I,r}, \tag{13}$$

$\widetilde{W}_{i,c_r} = \frac{w_r(d_{i,c_r})/\hat{v}_{i,c_r}^2}{\sum_{c_r=1}^{C_r} w_r(d_{i,c_r})/\hat{v}_{i,c_r}^2}$ and $\widetilde{w}_{I,c_r} = \frac{A_I V_{I,c_r}^{-1}}{\sum_I A_I^2 V_{I,c_r}^{-1}}$ are kernels that decay with $w_r(d_{i,c_r})$ and are scaled to sum to one. Therefore, when the bandwidth $h_r$ is sufficiently small such that the support of $w_r(d_{i,c_r})$ is entirely contained within the $I$-th aggregation unit, $\widetilde{W}_{i,c_r}$ and $\widetilde{w}_{I,c_r}$ both equal 1 in the $I$-th unit and 0 elsewhere, and thus Eq. (13) reduces to $\hat{z}_{i,r} = Z_{I,r}$. Even when this condition does not strictly hold, as the bandwidth decreases, Eq. (13) increasingly emphasizes the $I$-th aggregation unit through a weighted average of weighted averages, and $\hat{z}_{i,r}$ converges $Z_{I,r}$.

Suppose the following predicted value has been obtained:

$$\hat{y}_{i,R-1} = \sum_{k=1}^{K} a_i x_{i,k} \hat{\beta}_k + \hat{z}_{i,1:R-1}. \tag{14}$$

Its aggregated value is $\hat{Y}_{i,R-1} = \sum_{i \in I} \hat{y}_{i,R-1}$. To improve the predictive accuracy for $y_i$, we add the $R$-th single-scale process $\hat{z}_{i,R}$ on Eq. (14). Specifically, we fit $C_R$ aggregated local models (Eq. 10) on the residual $Y_i - \hat{Y}_{i,R-1}$ and synthesize the estimated models via Eq. (9). As with Eq. (13), this yields $\hat{z}_{i,R} = \sum_{c_r=1}^{C_r} \widetilde{W}_{i,c_r} \sum_{I=1}^{N} w_{I,c_r} \left(Y_i - \hat{Y}_{i,R-1}\right)$. Adding this to Eq. (14) gives:

$$\hat{y}_i = \hat{y}_{i,R-1} + \sum_{c_r=1}^{C_r} \widetilde{W}_{i,c_r} \sum_{I=1}^{N} \widetilde{w}_{I,c_r} \left(Y_I - \hat{Y}_{I,R-1}\right). \tag{15}$$

For the same reason as above, when the bandwidth is sufficiently small, the second term on the right-hand side can be approximated by $Y_I - \hat{Y}_{I,R-1}$, yielding $\hat{y}_i \approx \hat{y}_{i,R-1} + Y_I - \hat{Y}_{I,R-1}$. Aggregating both sides yields $\sum_{i \in I} \hat{y}_i \approx Y_I$, approximating the aggregation constraint (Eq. 1). Note that although here we assumed Eq. (14) for $\hat{y}_{i,R-1}$, this property holds for any $\hat{y}_{i,R-1}$.

When the local process $\hat{z}_{i,R}$ cannot accurately approximate the aggregation constraint, another finer-scale process $\hat{z}_{i,R+1}$ defined with a smaller bandwidth can be added. This inclusion provides an additional adjustment through Eq. (15). Repeating this procedure with progressively smaller bandwidths enables increasingly accurate approximation of the aggregation constraint.

In short, estimating $C_R$ local models with small-bandwidth kernels and synthesizing them via Eq. (9) yields a predictive process $\hat{z}_{i,1:R}$ that adjusts an arbitrary predictor $\hat{y}_{i,R-1}$ into an aggregation consistent predictor $\hat{y}_i$, and additional processes with smaller bandwidths can be incorporated to further improve the aggregation consistency. Building on this property, the next section proposes an algorithm that gradually decreases the bandwidth, learns the spatial pattern at each scale through local model synthesis, and sequentially adds $\hat{z}_{i,r}$ to the predictor. This learning process terminates once the bandwidth becomes sufficiently small and the aggregation constraint is confirmed to be approximately satisfied and the validation loss can no longer be improved.

## 3.4 Learning algorithm

This section describes the algorithm optimizing the number of scales *R*, which was assumed given in the previous section. Holdout validation (HV) is used for the optimization. In the HV, the aggregated data are randomly divided into an $\alpha\%$ of training set $Y_{I_t} \in \{Y_{1_t}, \dots, Y_{N_t}\}$ and $100(1-\alpha)\%$ of validation set $Y_{I_v} \in \{Y_{1_v}, \dots, Y_{N_v}\}$, where we set $\alpha = 0.75$. *R* is then optimized by minimizing the aggregated sum-of-squared error (SSE), which is consistent with Eq. (5):

$$SSE_R^{valid} = \sum_{I_v=1}^{N_v} \frac{1}{\sum_{i \in I_v} a_i^2} \left(Y_{I_v} - \hat{Y}_{I_v}\right)^2, \quad (16)$$

where $\hat{Y}_{I_v} = \sum_{i \in I_v} \sum_{k=1}^{K} a_i x_{i,k} \hat{\beta}_k + \sum_{r=1}^{R} b_r \hat{Z}_{I_v,r}$.

Figure 1 illustrates the training procedure. The learning process begins with a coarse-scale spatial process and then sequentially learns increasingly finer-scale spatial processes by progressively decreasing the bandwidth following $h_r = \alpha h_{r-1,}$. As discussed above, smaller bandwidths lead to stronger conformity with the aggregation constraint. Consequently, as the bandwidth decreases, there exists a scale, which is *r* in Figure 1, beyond which the aggregation constraint is approximately satisfied. The optimal scale is then selected as the scale *R* at or beyond *r* that minimizes $SSE_R^{valid}$. Through this procedure, CF-DS learns a spatial process that not only achieves low generalization error but also approximately satisfies the aggregation constraint.

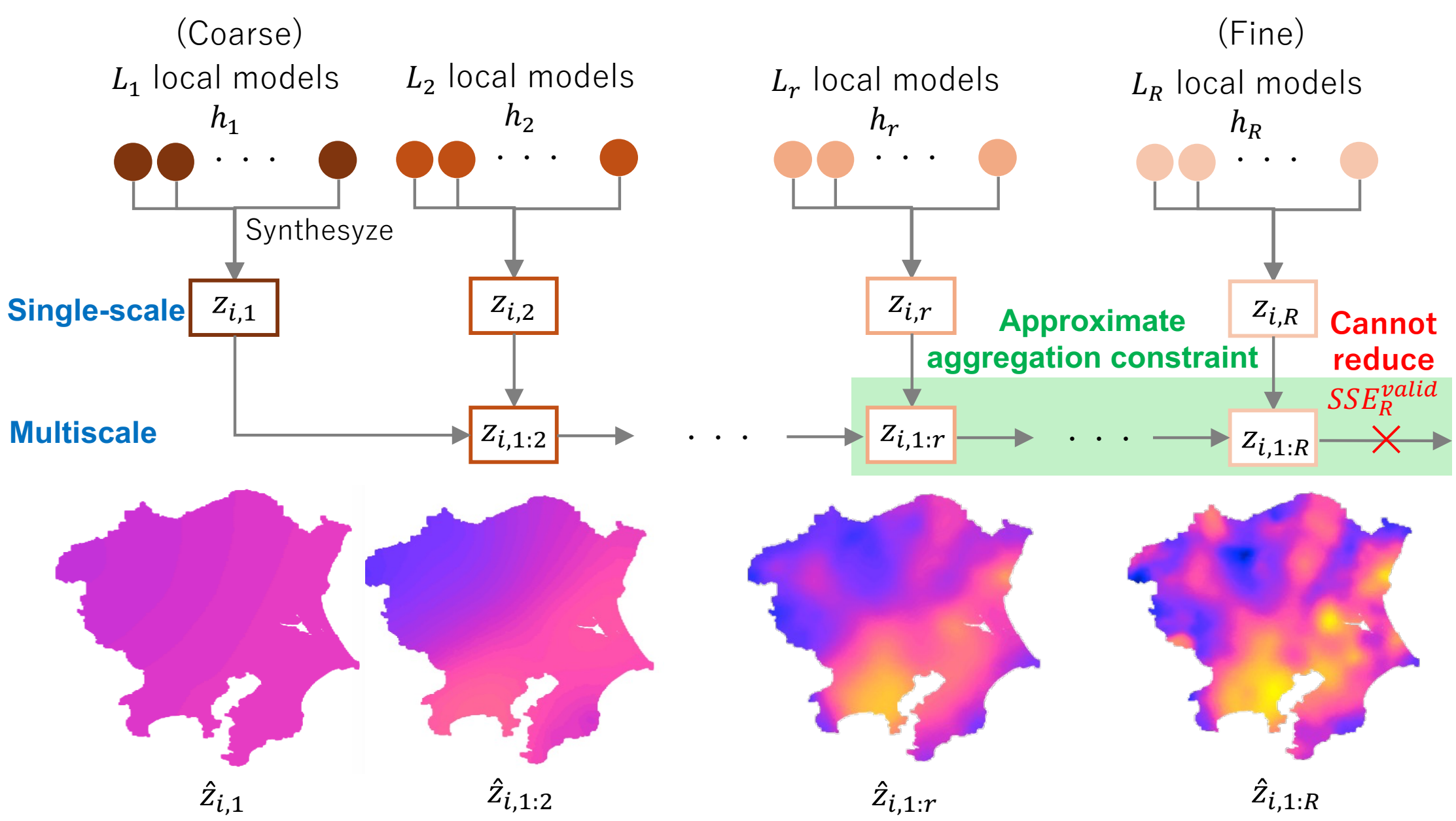

Figure 1: Learning procedure. As illustrated, local models are synthesized to construct a spatial process at each scale. These scale-wise processes are then sequentially added while progressively reducing the bandwidth, until the aggregation constraint is approximately satisfied and $SSE_R^{valid}$ is minimized.

The learning procedure is summarized as follows:

1. Initialize $R = 1$, $\hat{Z}_{I,1:R-1} = \hat{z}_{I,1:R-1} = 0$, $\hat{b}_{R-1} = 1$. Specify $\hat{\beta}_k$ by the weighted least squares estimator of $\beta_k$ minimizing $SSE_{R-1}^{train} = \sum_{I_t=1}^{N_t} \frac{1}{\sum_{i\in I_t} a_i^2} \left(Y_{I_t} - \sum_{k=1}^{K} \sum_{i\in I_t} a_i x_{i,k} \beta_k\right)^2$.
2. Estimate $\hat{Z}_{I,R}$ by minimizing the training SSE, $SSE_R^{train} = \sum_{I_t=1}^{N_t} \frac{1}{\sum_{i\in I_t} a_i^2} \left(Y_{I_t} - \sum_{k=1}^{K} \sum_{i\in I_t} a_i x_{i,k} \hat{\beta}_k - \hat{Z}_{I_t,1:R-1} - \hat{b}_R Z_{I_t,R}\right)^2$, following the original CFSM, as follows:

2.1. Distribute $C_R$ local centers over the study area. Similar to the original CFSM, we place the centers at the $k$-means cluster centroids, and assume $C_R = round(1.5, D^2/h_R^2)$, where $D$ is the diagonal length of the bounding box enclosing the sample sites.

2.2. For each center $c_R \in \{1_R, \dots, C_R\}$, fit the aggregated local model (Eq. 10) to estimate the predictive mean $E[z_{i,R}|c_R]$ and variance $V[z_{i,R}|c_R]$ of $z_{i,R}|c_R$ (Eq. 8).

2.3. Synthesize the estimated local predictive means $E[z_{i,R}|1_R], \dots, E[z_{i,R}|C_R]$ and variances $V[z_{i,R}|1_R], \dots, V[z_{i,R}|C_R]$ using Eq. (9) to form the disaggregated process $z_{i,R} \sim N(\hat{z}_{i,R}, \hat{\sigma}_{i,R}^2)$.

2.4. Evaluate the aggregated predictive mean $\hat{Z}_{I,R} = \sum_{i \in I} \hat{z}_{i,R}$.

3. Optimize $\beta_k$ and $b_R$ by minimizing $SSE_R^{train} = \sum_{I_t=1}^{N_t} \frac{1}{\sum_{i \in I_t} a_i^2} \left(Y_{I_t} - \sum_{k=1}^{K} \sum_{i \in I_t} a_i x_{i,k} \beta_k - \hat{Z}_{I_t,1:R-1} - b_R \hat{Z}_{I,R}\right)^2$ under the constraint that $0 \le b_R \le 1$. It is easily achieved by the constraint weighted least squares estimation using the training samples. These estimators are denoted as $\hat{\beta}_k$ and $\hat{b}_R$.

4. Evaluate the disaggregated process $\hat{z}_{I,1:R} = \hat{z}_{I,1:R-1} + \hat{b}_R \hat{z}_{I,R}$ and aggregated process $\hat{Z}_{I,1:R} = \hat{Z}_{I,1:R-1} + \hat{b}_R \hat{Z}_{I,R}$.

5. Evaluate $\hat{Y}_{I_t} = \sum_{k=1}^{K} \sum_{i \in I_t} a_i x_{i,k} \hat{\beta}_k + \hat{Z}_{I_t,1:R}$ and $\hat{Y}_{I_v} = \sum_{k=1}^{K} \sum_{i \in I_v} a_i x_{i,k} \hat{\beta}_k + \hat{Z}_{I_v,1:R}$.

6. Examine if the aggregated outputs $\hat{Y}_{I_t}$ accurately approximate the aggregation constraint. In this study, the accuracy is considered sufficient if (the 95-th percentile of $|Y_{I_t} - \hat{Y}_{I_t}|$) > $(0.1 \times$ Standard deviation of $Y_{I_t}$). If this condition is satisfied, proceed to Step 7. If not, an additional smaller bandwidth is required to satisfy the constraint, and so proceed to Step 8.

7. Evaluate $SSE_R^{valid}$. If $SSE_R^{valid} < \min(SSE_1^{valid}, \dots, SSE_{R-1}^{valid})$, $Q = 0$. Otherwise, $Q \to Q + 1$. If $Q$ is smaller than a threshold value, which is 5 in our case, proceed to Step 8. Otherwise, terminate the algorithm. The terminal resolution is $R$.

8. Update scale $R \to R + 1$, reduce the bandwidth $h_{R+1} = \delta h_R$, where we assumed $\delta = 0.9$, and go back to step 2.

After the sequential HV, the same procedure without steps 6 and 7 is repeated from $R$ = 1 to the selected $R$ using the entire samples to obtain the predictive value $\hat{y}_i^0 = \sum_{k=1}^{K} a_i x_{i,k} \hat{\beta}_k + \hat{z}_{i,1:R}$. Then, it is adjusted to satisfy the aggregation constraint exactly as $\hat{y}_i = \frac{Y_I}{\sum_{i \in I} \hat{y}_i^0} \hat{y}_i^0$. Alternatively, if the predictive values must be non-negative, the adjusted predictor is $\hat{y}_i = \frac{Y_I}{\sum_{i \in I} (\hat{y}_i^0)_+} (\hat{y}_i^0)_+$ where $(x)_+ = \max(x, 0)$.

Although we rely on a post-hoc adjustment in Step 9, the required adjustment is typically small as demonstrated later (see Figure 6) because $\hat{y}_i^0$ closely approximates the aggregation constraint a result of Step 6. If the aggregation constraint does not need to be satisfied exactly, for example, when the aggregated data are noisy, Step 9 can be omitted.

The estimation of $b_R$ in Step 3 is required to ensure that $SSE_R^{train} \leq SSE_{R-1}^{train}$ at each iteration. Specifically, $SSE_R^{train} = SSE_{R-1}^{train}$ if $b_R = 0$. Therefore, if $\hat{Z}_{I,R}$ cannot reduce the SSE, the estimation yields $\hat{b}_R = 0$ resulting in $SSE_R^{train} = SSE_{R-1}^{train}$. Otherwise, $\hat{b}_R > 0$ is estimated to reduce the training SSE. Consequently, Step 3 never increases the SSE. In addition, we assumed $b_R \leq 1$ to avoid potential overfitting.

The resulting predictor $\hat{y}_i$ minimizes the validation SSE while satisfying or approximating the aggregation constraint. In addition, because the method does not require likelihood evaluation or inversion of covariance matrix, it is computationally much faster than conventional spatial statistical methods, including ATP kriging. The next section examines its downscaling accuracy through Monte Carlo experiments.

# 4. Monte Carlo experiments

Section 4.1 compares the predictive accuracy of CF-DS with alternative methods, while Section 4.2 compares their computational efficiency especially for large size samples. Computations in Sections 4 and 5 were conducted using R 4.6.10, on a Mac Studio equipped with an Apple M3 Ultra chip and 512 GB of unified memory.

## 4.1 Predictive accuracy

We assume that the disaggregated units are $n \times n$ regular grid points spaced at intervals of 1.0, and that the aggregated units are obtained by aggregating them into $N \times N$ grids where $N = \frac{n}{4}$; that is, each aggregated unit consists of $4 \times 4$ disaggregated units. The task is to predict the disaggregated values $y_i$ from the aggregated observations $Y_I$. The aggregated data are generated by the following model:

$$Y_I = \sum_{i \in I} y_i, \qquad y_i = a_i \left( \sum_{k=1}^{3} x_{ik}\beta_k + \sum_{j=1}^{N} \tilde{w}_h(d_{ij})\, u_j + e_i \right)_+, \tag{17}$$
$$a_i \sim Unif(0,1), \qquad u_j \sim N(0,1), \qquad e_j \sim N(0,1),$$

where $\beta_2 = 2.0$, $\beta_3 = -0.5$. $\tilde{w}_h(d_{ij}) = w_h(d_{ij}) / \sum_j w_h(d_{ij})$ is a row-standardized weight, where $w_h(d_{ij}) = \exp(d_{ij}/h)$ with $h = 3$. The covariates are assumed to be spatially dependent, as follows:

$$x_{ik} = 0.5 \sum_{j=1}^{N} \tilde{w}_h(d_{ij})\, u_{jk} + 0.5 e_{ik}, \qquad u_{jk} \sim N(0,1), \qquad e_{ik} \sim N(0,1). \tag{18}$$

For each case, with $\beta_1 \in \{0.5, -1.5\}$ and $n \in \{20^2, 40^2, 60^2\}$ corresponding to $N \in \{25, 100, 225\}$, downscaling is performed 200 times. The setting $\beta_1 = 0.5$ yields few zero values, whereas $\beta_1 = -1.5$ yields many zeros. We compared the RMSE of the predictive value $\hat{y}_i$ obtained from following models: areal weighting interpolation (Areal), dasymetric mapping (Dasymetric), which distributes data in proportion to $a_i$, and area-to-point kriging (ATP kriging), which treats $a_i$ as an proportional weight, and $x_{i1}, x_{i2}, x_{i3}$ as covariates, and the proposed CF-DS, which uses the same variables. CF-DS is implemented using Rcpp (https://cran.r-project.org/web/packages/Rcpp/index.html) to enable scalable computation.

Figure 2 reports RMSE values. In case with small sample sizes of $N = 25$, ATP kriging and CF-DS were less stable than Dasymetric, and no clear improvement in accuracy was observed. By contrast, when $N \geq 100$, the predictive accuracy of ATP kriging and CF-DS was superior to that of Areal and Dasymetric. These results demonstrate that spatial process-based methods improve downscaling accuracy when $N$ is sufficiently large. Moreover, CF-DS achieved accuracy comparable to ATP kriging across all cases, suggesting that CF-DS can serve as a viable alternative to ATP kriging.

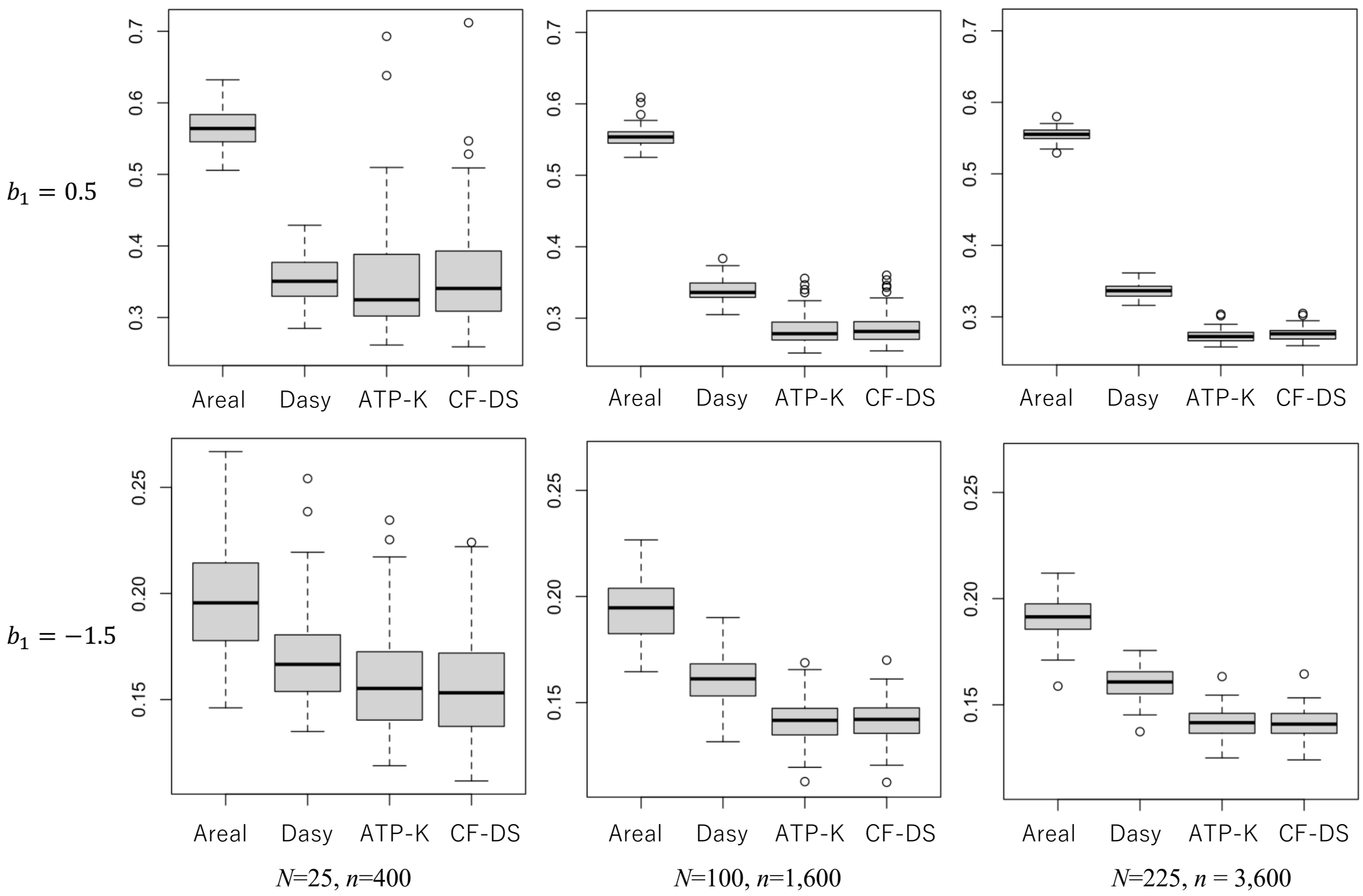

Figure 2: Boxplots of the RMSEs for extensive data. Dasy and ATP-K denote dasymetric mapping and ATP kriging, respectively.

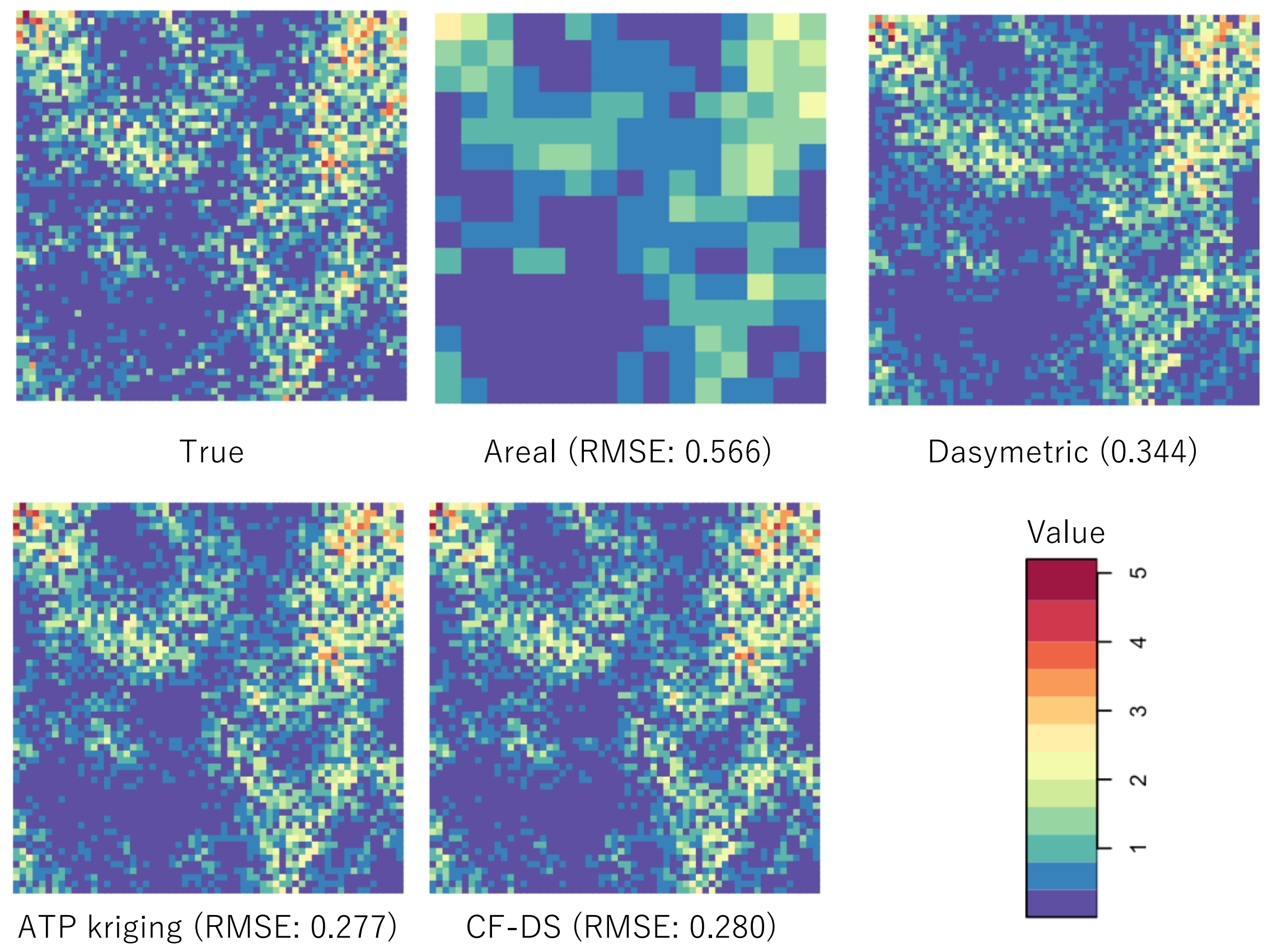

Figure 3: Downscaling result in an iteration ($n = 60^2$; $b_1 = 0.5$)

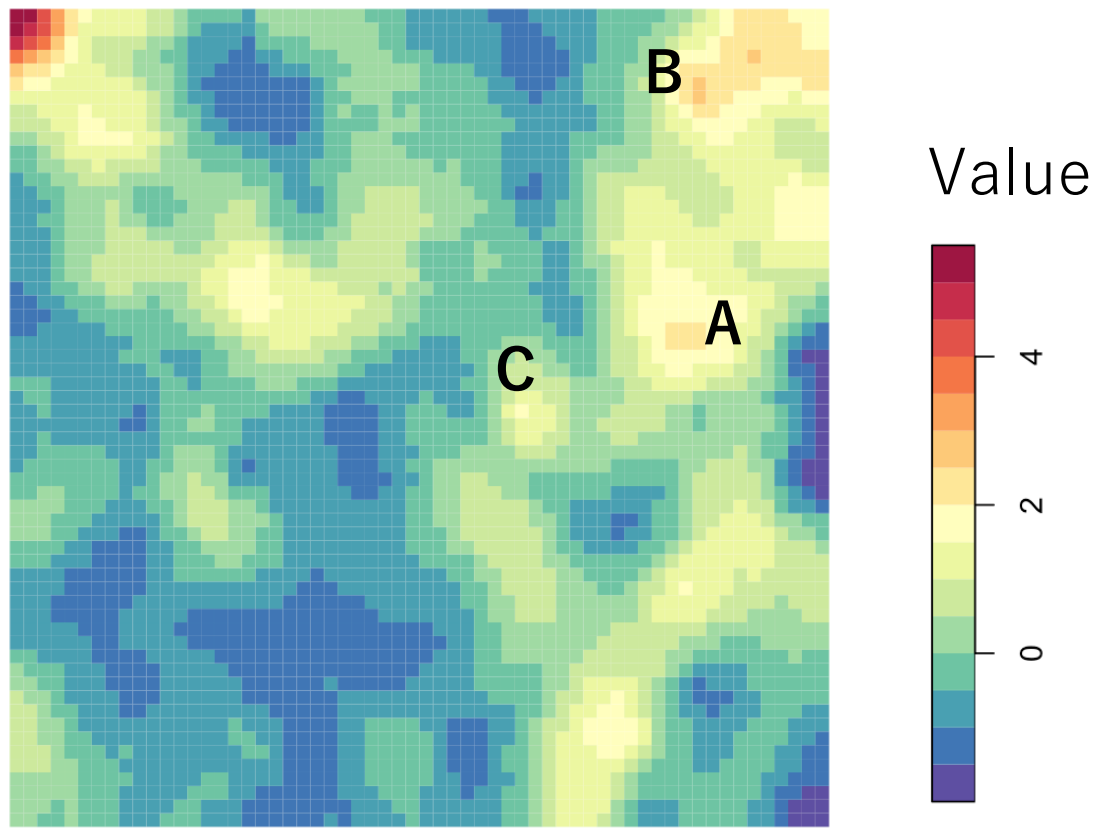

Figure 4: Predictive mean $\hat{z}_{i,1:R}$ of the spatial process obtained from CF-DS ($n = 60^2$; $b_1 = 0.5$)

Figure 3 shows an example of the downscaling results produced by each method. The result of Areal, which does not use auxiliary variables, was poor, confirming the importance of incorporating auxiliary variables in downscaling. In contrast, the other methods produced patterns generally similar to the true values, demonstrating their high level of predictive accuracy. The results of CF-DS were similar to those of Dasymetric and ATP kriging. Figure 4 shows the predictive mean $\hat{z}_{i,1:R}$ of the spatial process estimated by CF-DS. The value range $\hat{z}_{i,1:R}$ is comparable to that of the predictive values $\hat{y}_i$ themselves, implying that strong spatially dependent patterns were successfully captured. As a result in regions A - C shown in Figure 4, the predictive values obtained from CF-DS are larger than Dasymetric (see Figure 3), suggesting that CF-DS locally adjusted the predictive value based on the latent spatial process.

For reference, we also conducted an additional experiment assuming $a_i = 1$ in the data-generating process (Eq. 17), representing a situation where no informative allocation variable exists; this scenario may be likely, for example, in meteorological downscaling. The downscaling result is shown in Figure 5. In this case, the proportional weights used by Dasymetric become $a_i = 1$, making it equivalent to Areal. As a result, Dasymetric failed to capture fine-scale patterns in this case. In contrast, ATP kriging and CF-DS accurately captured the disaggregated-level patterns, demonstrating their ability to perform effective downscaling even in the absence of informative allocation variables.

Similar results were also obtained for the downscaling of intensive data (see Section 2). See

Appendix. 2 for further details.

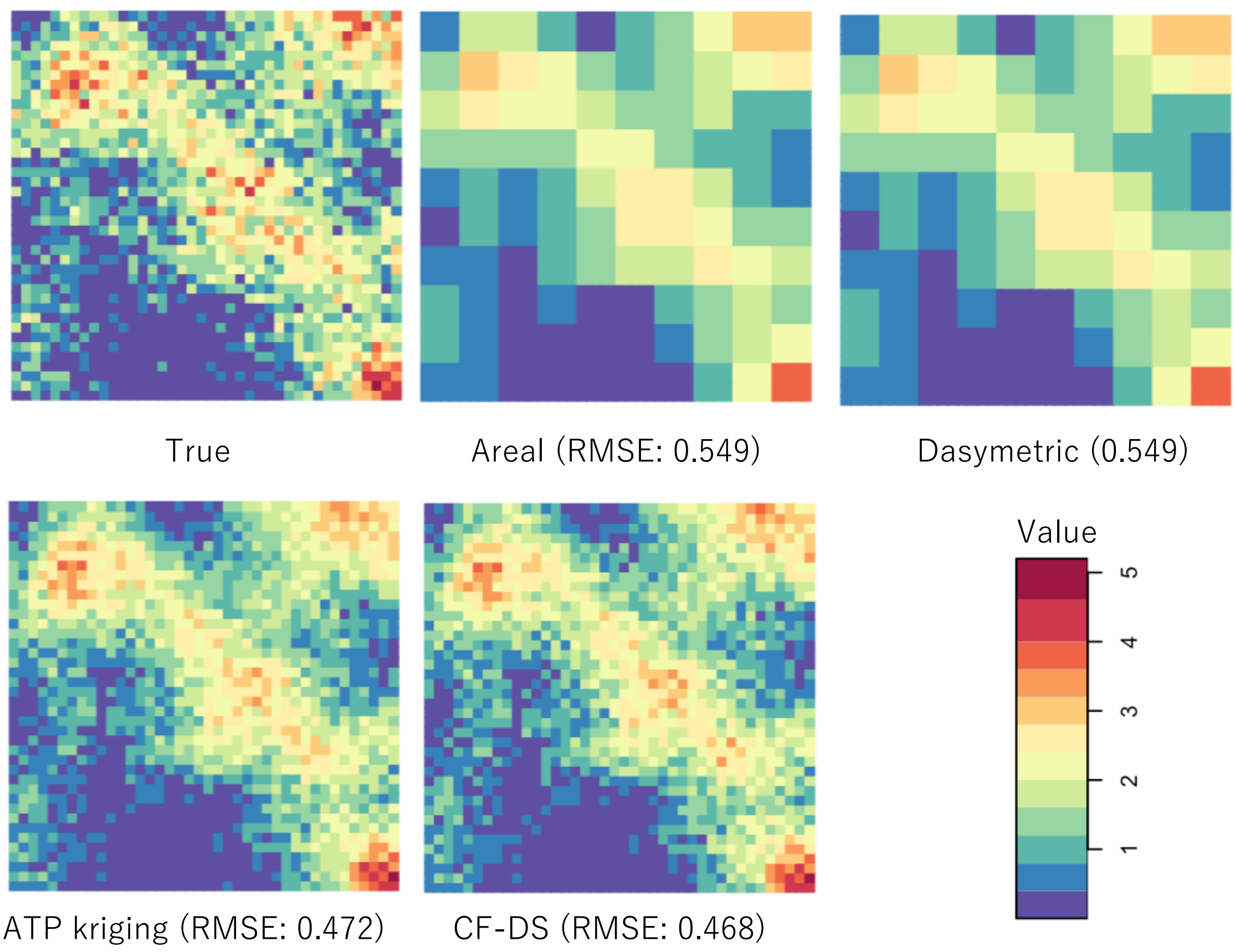

Figure 5: Downscaling results in case without an offset variable ($n = 40^2$; $b_1 = 0.5$)

## 4.2 Computational efficiency

Under the same problem setting as in Section 4.1 with $b_1 = 0.5$, we compare the computation time of ATP kriging and CF-DS. Looking ahead to large-size applications, we considered the numbers of disaggregated units $n \in \{400,1600,3600,6400,14400,25600,40000\}$ and corresponding the numbers of aggregated units $N \in \{25,100,225,400,900,1600,2500\}$. For each case, we ran 10 replications and evaluated the median computation time.

The results are summarized in Table 1. The computation time of ATP kriging increases sharply with increase of the sample size, whereas that of CF-DS grows only linearly. For example, when $N = 225$ and $n = 3{,}600$, ATP kriging required 217.13 seconds, while CF-DS required only 1.04 seconds. Given that the two methods achieved a comparable level of predictive accuracy, this is a remarkable improvement. Even in the heaviest case with $N = 2{,}500$ and $n = 40{,}000$, where ATP Kriging is difficult to apply, CF-DS required only 23.41 seconds. These results suggest that CF-DS is particularly useful for large-size downscaling problems.

Table 1: Median computation time (second)

| $n$ | $N$ | ATP kriging | CF-DS |
|---:|---:|---:|---:|
| 400 | 25 | 0.59 | 0.11 |
| 1,600 | 100 | 16.16 | 0.45 |
| 3,600 | 225 | 217.13 | 1.04 |
| 6,400 | 400 | NA | 2.13 |
| 14,400 | 900 | NA | 5.73 |
| 25,600 | 1,600 | NA | 12.07 |
| 40,000 | 2,500 | NA | 23.41 |

# 5. Application

## 5.1 Outline

Understanding electricity consumption at fine spatial units is essential for efficient energy management, including the effective integration of renewable energy sources, demand-response strategies, and resilient power systems (Jordehi, 2019; Mahmood et al., 2024). Accordingly, this study applies CF-DS and dasymetric mapping to downscale electricity consumption in the commercial sector (source: Japan Energy Database; https://energy-sustainability.jp/) from municipalities ($N = 308$; Figure 6) to 1-km grid cells ($n = 31,172$) in the Tokyo metropolitan area.[4]

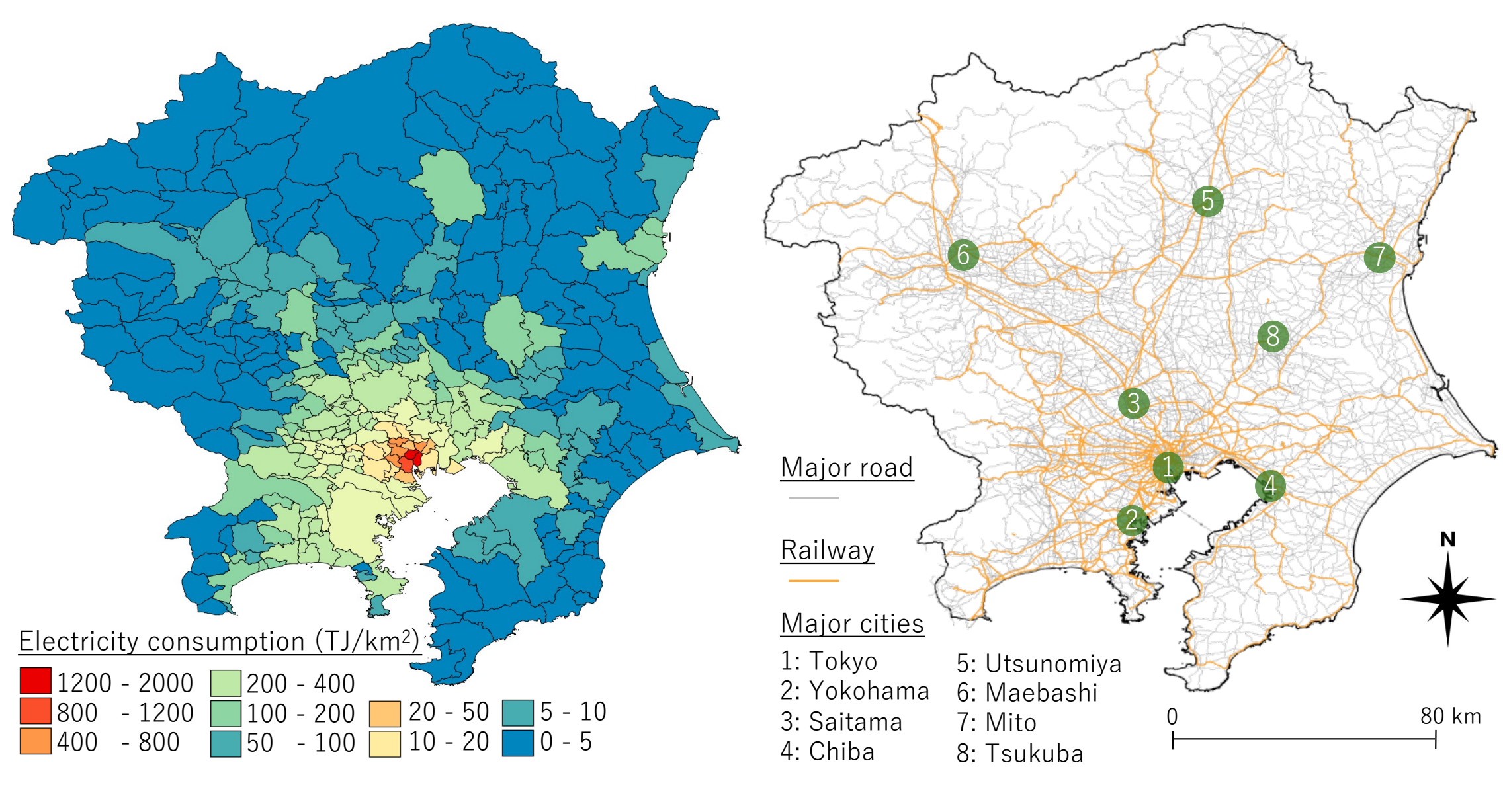

Figure 6: Municipal commercial electricity consumption (left) and major cities with transportation networks in the study area (right). The cities 1- 7 are prefectural capitals.

[4] ATP kriging could not be estimated, mainly because the number of disaggregated units was too large.

Dasymetric mapping distributes the municipal electricity use in proportion to the total floor area of the tertiary industry (TFA_Office). Similarly, CF-DS considers the TFA_Office as the proportional weight $a_i$, while additionally accounting for covariates, including the number of tertiary industry establishments (Num_Office), and proximity to train station (Train), major road (Road), and an indicator or commercial-use district (Commerce), listed in Table 2. As mapped in Figure 6 (right), both the railway and road networks are highly developed across the study area. In particular, railways are widely used for commuting, and major cities are distributed along the network. Therefore, electricity consumption is expected to be high around railway stations.

Table 2: Covariates and offset variable

| Category | Name | Description | Source |
|---|---|---|---|
| Covariates | Num_Office | Number of tertiary-industry establishments | Economic [1)] Census |
| | Train | Proximity to train station, defined by $\exp(-[\text{distance to the nearest railway station}]^2)$ | NLNI |
| | Road | Proximity to major roads defined by $\exp(-[\text{distance to the nearest major road}]^2)$ | OpenStreetMap |
| | Commerce | Proximity to commercial-use districts, defined as $\exp(-[\text{distance to the nearest commercial-use district}]^2)$ | NLNI |
| Offset variable | TFA_Office | Total floor area of the tertiary industry | ZENRIN CO., LTD. |

[1)] Economic Census (https://www.stat.go.jp/english/data/e-census.html), National Land Numerical Information download site (https://nlftp.mlit.go.jp/ksj/), OpenStreetMap (https://www.openstreetmap.org/)

## 5.2 Result

The number of scales optimized during the CF-DS training is $R = 41$. The resulting spatial process $z_{1:R}$ comprises 41 single-scale processes with bandwidths ranging from 111 km to 1.64 km, suggesting that electricity consumption exhibits multiscale spatial variations. The downscaling results are shown in Figure 7. Compared with dasymetric mapping, CF-DS estimates higher electricity consumption along railway lines. This result is intuitively reasonable, as the study area is characterized as a train-oriented metropolis in which urban development has been concentrated along the railway network (Yamagata and Seya, 2013). Dasymetric mapping produces a more dispersed spatial pattern. Given the superior predictive accuracy of CF-DS demonstrated in Section 4, this pattern may be an artifact arising from the strong assumption in dasymetric mapping that electricity consumption is strictly proportional to TFA_Office.

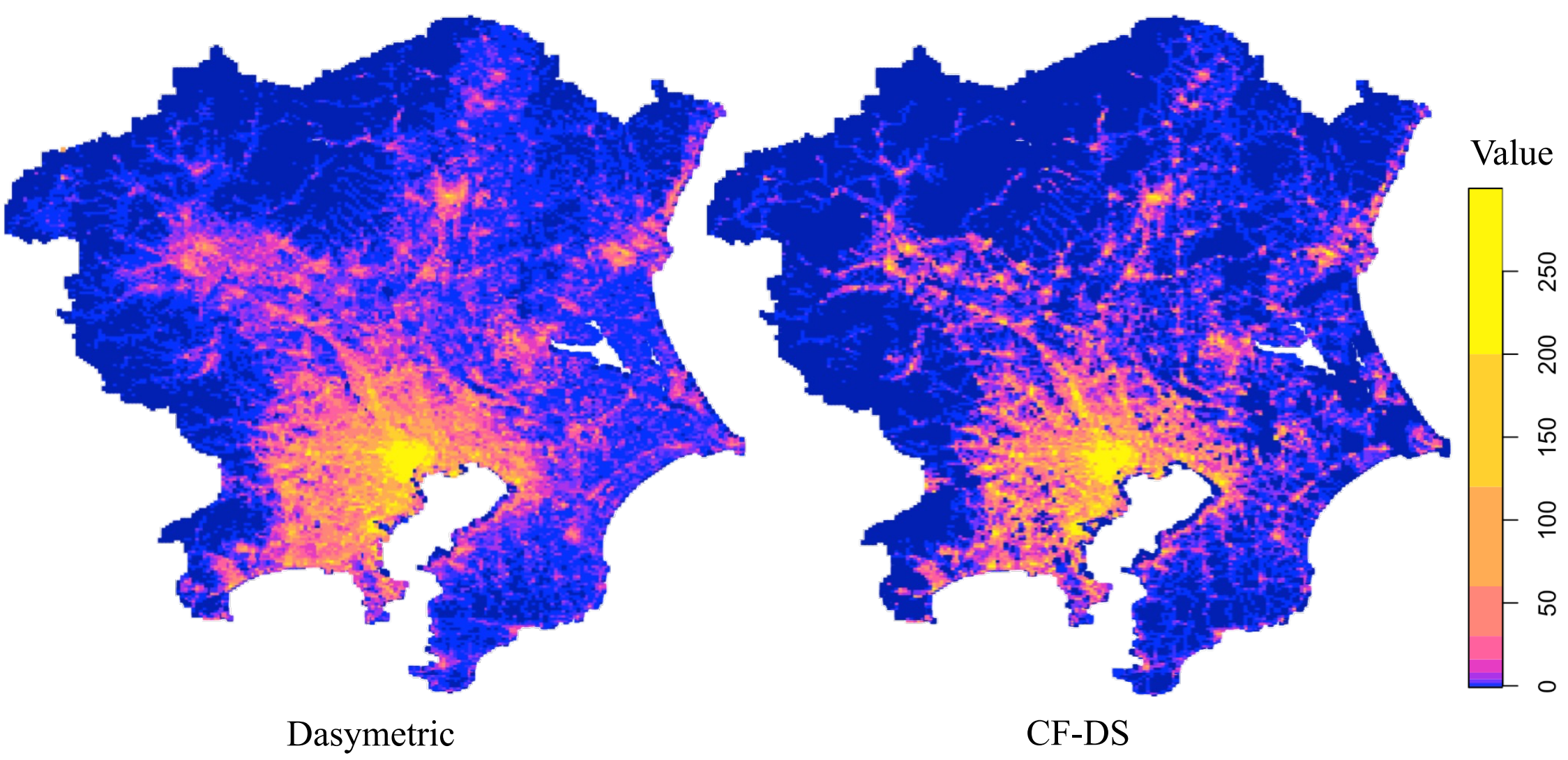

Figure 7: Downscaling results

While CF-DS adjusted the predictor to exactly satisfy the aggregation constraint (Step 9 of the learning algorithm. See Section 3.4), Figure 8 compares the aggregated predictions before and after the adjustment. The difference between these two is small, indicating that only a minor adjustment was made in Step 9 of the algorithm. Furthermore, even without the adjustment, the aggregated predictions were generally consistent with the observed values, confirming that our local modeling approach successfully approximates the aggregation constraint.

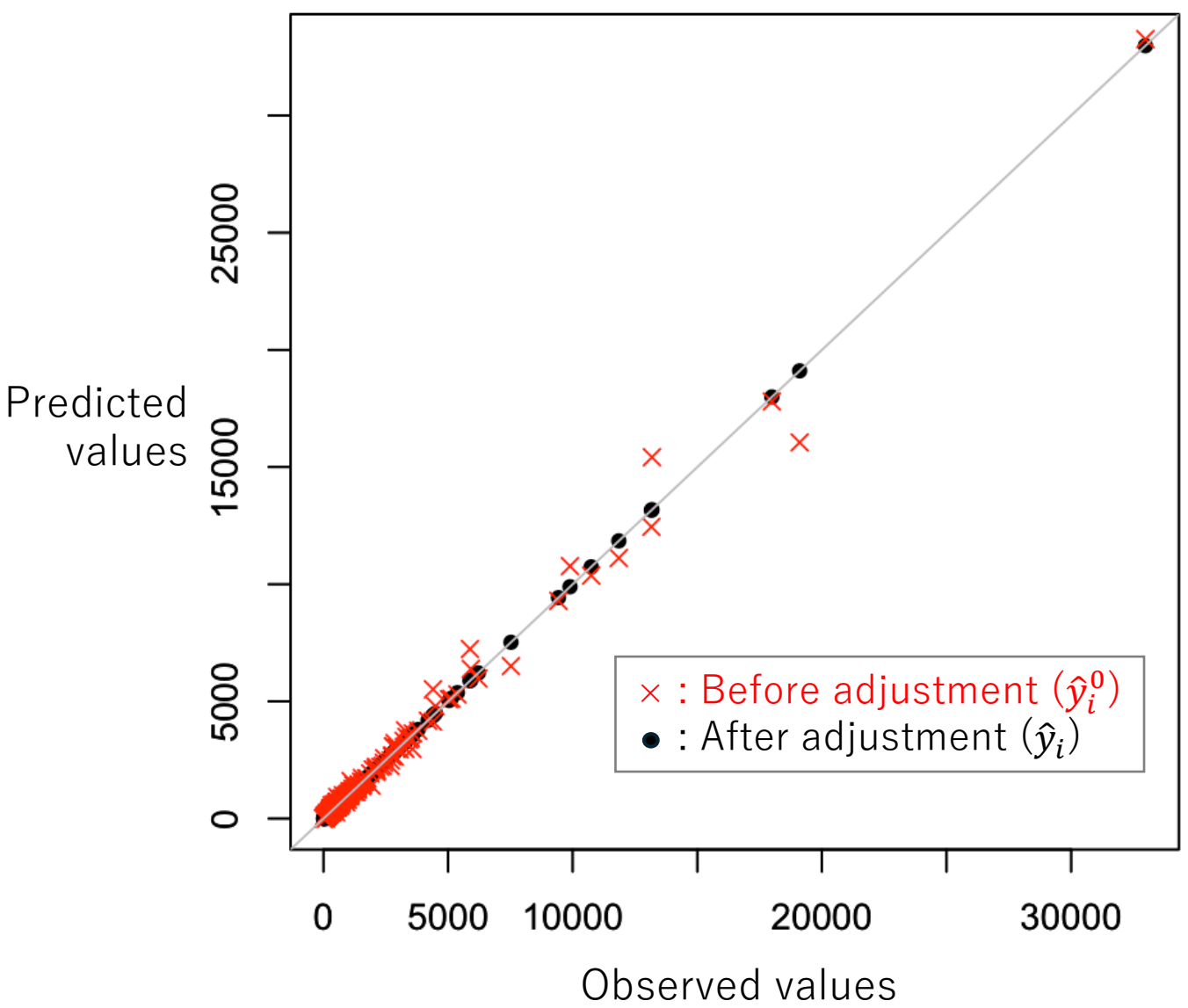

Figure 8: Observed and predictive values at the aggregated level before and after the adjustment

Table 3 summarizes the regression coefficients estimated from CF-DS. Because TFA_Office is used as an offset variable, each coefficient represents the effect of the interaction between TFA_Office and the covariate. For comparability, the reported values are scaled to correspond to coefficients on standardized interaction variables with unit variance. The coefficient on TFA_Office itself is negative, while those on the interaction variables are all positive. This suggests that the interaction variables have greater positive associations with electricity consumption than TFA_Office itself. In particular, the coefficient on TFA_Office × Pot_Road was the largest, indicating the strongest explanatory contribution. When a suitable variable for the proportional distribution is available, dasymetric mapping is known to achieve high predictive accuracy (Eicher and Brewer, 2001). The strength of CF-DS lies in its ability to improve predictive accuracy by leveraging covariates and the latent spatial pattern even when such a variable is unavailable.

Table 3: Estimated coefficients on the scaled covariates.

| Offset variable × Covariates | | Estimate |
|---|---|---|
| | 1 (Intercept) | −263.45 |
| | Num_Office | 33.36 |
| TFA_Office × | Pot_Rail | 20.92 |
| | Pot_Road | 267.32 |
| | Pot_Commerce | 19.85 |

Figure 9 decomposes the predictive values of CF-DS into the components explained by the covariates and those explained by the spatial process. The above-discussed concentration around train stations is picked up by the former. The spatial process, in turn, captures localized increases in electricity consumption in the areas surrounding central Tokyo, including Yokohama, Saitama, and Chiba, as well as outer suburban cities such as Tsukuba, Utsunomiya, and Mito (see Figure 6 (right)). Figure 9 suggests that both the covariates and the spatial process adjust the TFA_Office-based baseline prediction in an intuitively reasonable manner.

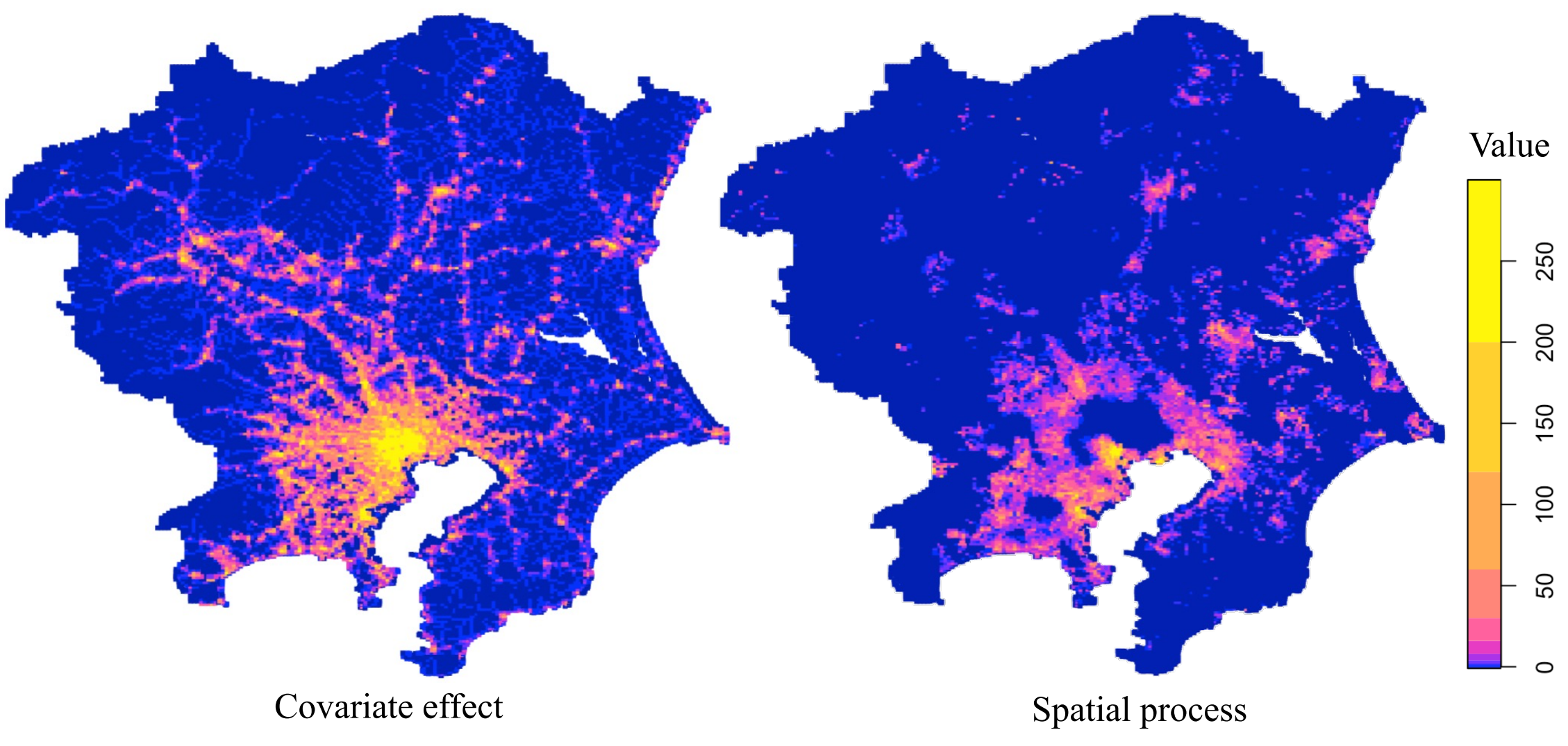

Figure 9: Estimated covariate effects and spatial processes in the CF-DS predictions

# 6. Concluding remarks

This study proposes coarse-to-fine downscaling (CF-DS), a novel downscaling method built upon the coarse-to-fine spatial modeling (CFSM) framework. Conventional spatial process-based downscaling methods reply on covariance modeling, which is computationally demanding and has limited their applicability to large-size problems. In contrast, CF-DS learns the spatial process without explicitly modeling covariance, thereby substantially reducing the computational cost. Monte Carlo experiments show that CF-DS achieves predictive accuracy comparable to ATP kriging while dramatically reducing computation time. The application of electricity consumption downscaling further illustrates its practical utility. In particular, CF-DS successfully identifies elevated electricity consumption in urban areas including neighborhoods surrounding railway stations that could not be captured by dasymetric mapping.

In recent years, large-size downscaling, including global-scale applications has attracted increasing attention in socio-economic domain and related domains (see Leyk et al., 2019; Lei et al., 2023; for review). However, the computational burden of conventional spatial-statistical methods has often limited their applicability to such problems. CF-DS addresses this limitation and provides one of the few practical approaches capable of performing a large-size downscaling problem while accounting for underlying spatial patterns.

Several issues remain for future work. First, CF-DS should be compared with existing scalable DS methods. Recent approaches include those based on the stochastic partial differential equation (SPDE) method (e.g., Rodriguez Avellaneda et al., 2026), random forests (e.g., Stevens et al., 2015), and neural networks (e.g., Ulyanov et al., 2017). Second, CF-DS can be extended to cases where the aggregation constraint should not be satisfied exactly. Such situations include, for example, the case with noisy aggregated data. Because the main algorithm in CF-DS only approximates the aggregation constraint, it has the potential to yield accurate predictions even in such cases, in contrast to conventional DS methods, which impose the aggregation constraint as a hard constraint. Third, CF-DS can be extended to spatiotemporal data. To achieve this, CF-DS needs to be extended so that it can describe disaggregated spatiotemporal processes from aggregated data whose aggregation units may change over time due to, for example, municipal mergers or national consolidations. Fourth, the method can be extended to address a broader range of COSPs. In many applications, data observed over incompatible spatial supports, such as points, grid cells, and municipal units, must be analyzed jointly (e.g., Wang et al., 2021). Extending CF-DS for such analysis while maintaining computational efficiency, would substantially broaden its potential applications.

## Data availability

CF-DS is implemented in an R package spCF (https://cran.r-project.org/web/packages/spCF/index.html).

# References

- Atkinson, P. M. (2013). Downscaling in remote sensing. *International Journal of Applied Earth Observation and Geoinformation*, 22, 106–114.
- Cao, Y., & Fleet, D. J. (2015). Transductive log opinion pool of Gaussian process experts. *ArXiv* 1511.07551.
- Chen, C., He, Q., & Li, Y. (2024). Downscaling and merging multiple satellite precipitation products and gauge observations using random forest with the incorporation of spatial autocorrelation. *Journal of Hydrology*, 632, 130919.
- Eicher, C. L., & Brewer, C. A. (2001). Dasymetric mapping and areal interpolation: Implementation and evaluation. *Cartography and Geographic Information Science*, 28(2), 125–138.
- Fisher, P. F., & Langford, M. (1996). Modelling sensitivity to accuracy in classified imagery: A study of areal interpolation. *The Professional Geographer*, 48(3), 299–309.
- Flowerdew, R., & Green, M. (1994). Areal interpolation and types of data. *Spatial analysis and GIS*, 121, 145.
- Forlani, C., Bhatt, S., Cameletti, M., Krainski, E., & Blangiardo, M. (2020). A joint Bayesian space–time model to integrate spatially misaligned air pollution data in R-INLA. *Environmetrics*, 31(8), e2644.

- Gelfand, A. E., & Schliep, E. M. (2025). Model-based spatial data fusion. *Annual Review of Statistics and Its Application*, 13.
- Gotway, C. A., & Young, L. J. (2002). Combining incompatible spatial data. *Journal of the American Statistical Association*, 97(458), 632–648.
- Kyriakidis, P. C. (2004). A geostatistical framework for area-to-point spatial interpolation. *Geographical Analysis*, 36(3), 259–289.
- Lei, Z., Xie, Y., Cheng, P., & Yang, H. (2023). From auxiliary data to research prospects, a review of gridded population mapping. Transactions in GIS, 27(1), 3-39.
- Leyk, S., Gaughan, A. E., Adamo, S. B., De Sherbinin, A., Balk, D., Freire, S., et al. (2019). The spatial allocation of population: a review of large-scale gridded population data products and their fitness for use. *Earth System Science Data*, 11(3), 1385–1409.
- Mahmood, M., Chowdhury, P., Yeassin, R., Hasan, M., Ahmad, T., & Chowdhury, N.-U.-R. (2024). Impacts of digitalization on smart grids, renewable energy, and demand response: An updated review of current applications. *Energy Conversion and Management: X*, 24, 100790.
- Moraga, P., & Alahmadi, H. (2026). Bayesian spatial disaggregation modeling for the detection of disease clusters. *Spatial Statistics*, 100982.

- Mugglin, A. S., Carlin, B. P., Zhu, L., & Conlon, E. (1999). Bayesian areal interpolation, estimation, and smoothing: An inferential approach for geographic information systems. *Environment and Planning A*, 31(8), 1337–1352.
- Murakami, D., Comber, A., Yoshida, T., Tsutsumida, N., Brunsdon, C., & Nakaya, T. (2026a). Coarse-to-fine spatial modeling: A scalable, machine-learning-compatible framework. *Geographical Analysis*, 58(2), e70034.
- Murakami, D., Comber, A., Yoshida, T., Tsutsumida, N., Brunsdon, C., & Nakaya, T. (2026b). Coarse-to-fine spatial GLMM for scalable prediction and multiscale analysis. *ArXiv*, 2605.01157.
- Rodríguez Avellaneda, F., Chacón-Montalván, E. A., & Moraga, P. (2026). Multivariate disaggregation modeling of air pollutants: A case-study of PM2.5, PM10 and ozone prediction in Portugal and Italy. *The American Statistician*, 80(1), 109–134.
- Sadahiro, Y. (2000). Accuracy of count data estimated by the point-in-polygon method. *Geographical Analysis*, 32(1), 64–89.
- Stengel, K., Glaws, A., Hettinger, D., & King, R. N. (2020). Adversarial super-resolution of climatological wind and solar data. *Proceedings of the National Academy of Sciences of the United States of America*, 117(29), 16805–16815.

- Stevens, F. R., Gaughan, A. E., Linard, C., & Tatem, A. J. (2015). Disaggregating census data for population mapping using random forests with remotely-sensed and ancillary data. *PLOS ONE*, 10(2), e0107042.
- Tomari, M., Seya, H., Murakami, D., Yamagata, Y., & Oki, T. (2025). Global gridded population projection via spatial econometric model considering rank-size rule. *Environmental Research Communications*, 7(10), 101002.
- Ulyanov, D., Vedaldi, A., & Lempitsky, V. (2018). Deep image prior. In *Proceedings of the IEEE Conference on Computer Vision and Pattern Recognition*.
- Vandal, T., Kodra, E., Ganguly, S., Michaelis, A., Nemani, R., & Ganguly, A. R. (2017). DeepSD: Generating high resolution climate change projections through single image super-resolution. In *Proceedings of the 23rd ACM SIGKDD International Conference on Knowledge Discovery and Data Mining* (pp. 1663–1672).
- Wang, Q., Shi, W., Atkinson, P. M., & Zhao, Y. (2015). Downscaling MODIS images with area-to-point regression kriging. *Remote Sensing of Environment*, 166, 191–204.
- Wang, S., Li, R., Jiang, J., & Meng, Y. (2021). Fine-scale population estimation based on building classifications: A case study in Wuhan. *Future Internet*, 13(10), 251.
- Yamagata, Y., & Seya, H. (2013). Simulating a future smart city: An integrated land use-energy model. *Applied Energy*, 112, 1466–1474.

# Appendix 1. Formulation for intensive data

For intensive data, the aggregation constraint is given by $Y_I^{(intens)} = \sum_{i\in I} \frac{a_i}{A_I} y_i^{(intens)}$ where $A_I = \sum_{i\in I} a_i$ (Eq. 2). By introducing a normalized weight $\tilde{a}_i = a_i/A_I$, which satisfies $\sum_{i\in I} \tilde{a}_i = 1$, the derivation in Section 3 carries over after by replacing $a_i$ with $\tilde{a}_i$. Accordingly, only the equations listed below need to be modified.

The weight $a_i$, which appears in the mean and variance terms in Eq. (4), drops out because $y_i^{(intens)} = \frac{y_i}{a_i}$ is already measured on a per-unit scale. The resulting disaggregated data model is given by:

$$y_i^{(intens)} = \sum_{k=1}^{K} x_{i,k}\beta_k + z_{i,1:R} + e_i, \quad e_i \sim N(0, \sigma^2). \tag{A1}$$

Aggregating Eq. (A1) yields the following aggregated model:

$$Y_I^{(intens)} = \sum_{i\in I}\sum_{k=1}^{K} \tilde{a}_i x_{i,k}\beta_k + Z_{I,1:R}^{(intens)} + E_I^{(intens)}, \quad E_I^{(intens)} \sim N\left(0, \sigma^2 \frac{\sum_{i\in I} a_i^2}{A_I^2}\right), \tag{A2}$$

where $Z_{I,1:R}^{(intens)} = \sum_{i\in I} \tilde{a}_i z_{i,1:R}$ is the weighted average of the process over area *I*.

Analogically, the local model (Eq. 7) no longer includes the weight $a_i$:

$$z_{i,r}|c_r = \mu_{c_r} + e_{i,c_r}, \quad e_{i,c_r} \sim N\left(0, \frac{v_{c_r}^2}{w_{h_r}^2(d_{i,c_r})}\right). \tag{A3}$$

The corresponding aggregated local model is given by:

$$Z_{I,r}^{(intens)}|c_r = \mu_{c_r} + E_{I,c_r}, \quad E_{I,c_r} \sim N\left(0, v_{c_r}^2 V_{I,c_r}^{(intens)}\right), \tag{A4}$$

where $V_{I,c_r}^{(intens)} = \sum_{i\in I} \frac{\tilde{a}_i^2}{w_{h_r}^2(d_{i,c_r})}$ denotes the aggregation variance for intensive data. The resulting

WLS estimators are then obtained as follows:

$$\hat{\mu}_{c_r} = \frac{\sum_I \left[V_{I,c_r}^{(intens)}\right]^{-1} Z_{I,r}^{(intens)}}{\sum_I \left[V_{I,c_r}^{(intens)}\right]^{-1}}, \qquad \hat{v}_{c_r}^2 = \frac{1}{N_{c_r} - 1} \sum_{I=1}^{N} \frac{\left(Z_{I,r}^{(intens)} - \hat{\mu}_{c_r}\right)^2}{V_{I,c_r}^{(intens)}}. \tag{A5}$$

Thus, by replacing Eqs. (4), (5), (7), (10), and (11) with Eqs. (A1) – (A5), the optimal $R$ is evaluated through the sequential HV procedure. Consistent with the variance in Eq. (A2), the validation SSE minimized in the HV is given by Eq. (A6).

$$SSE_R^{valid} = \sum_{I_v=1}^{N_v} \frac{A_{I_v}^2}{\sum_{i \in I} \tilde{a}_i^2} \left(Y_{I_v}^{(intens)} - \tilde{Y}_{I_v}^{(intens)}\right)^2 \tag{A6}$$

After optimizing $R$, the predictive value $\hat{y}_i^{0(intens)} = \sum_{k=1}^{K} x_{i,k} \hat{\beta}_k + \hat{z}_{i,1:R}$ is obtained. The post-hoc adjustment to exactly satisfy the aggregation constraint, corresponding to Step 9 of Section 3.4 for extensive data, is given by.

$$\hat{y}_i^{(intens)} = \frac{Y_I^{(intens)}}{\sum_{i \in I} \tilde{a}_i \hat{y}_i^{0(intens)}} \hat{y}_i^{0(intens)}. \tag{A7}$$

# Appendix 2. Downscaling accuracy for intensive data

This appendix compares the downscaling accuracy of Areal, Dasymetric, ATP kriging, and CF-DS for intensive variables. Similar to Section 4, the task is to predict the disaggregated intensity values $y_i$ from the aggregated intensity values $Y_I$. The aggregated data are generated according to the following model:

$$Y_I = \sum_{i \in I} \frac{a_i}{\sum_{i \in I} a_i} y_i^{(intens)}, \qquad y_i^{(intens)} = \left( \sum_{k=1}^{3} x_{ik}\beta_k + \sum_{j=1}^{N} \tilde{w}_h(d_{ij})\, u_j + e_i \right)_+, \tag{A8}$$

$$a_i \sim Unif(0,1), \qquad u_j \sim N(0,1), \qquad e_j \sim N(0,1).$$

The remaining simulation settings are identical to those in Section 4.

Figure A1 compares downscale accuracy of each method in terms of RMSE. The accuracy of CF-DS is almost identical to that of ATP kriging across all cases, and both methods achieve lower RMSE as *N* and *n* increase. This result is consistent with the extensive data case shown in Figure 2, confirming the accuracy of CF-DS for intensive data.

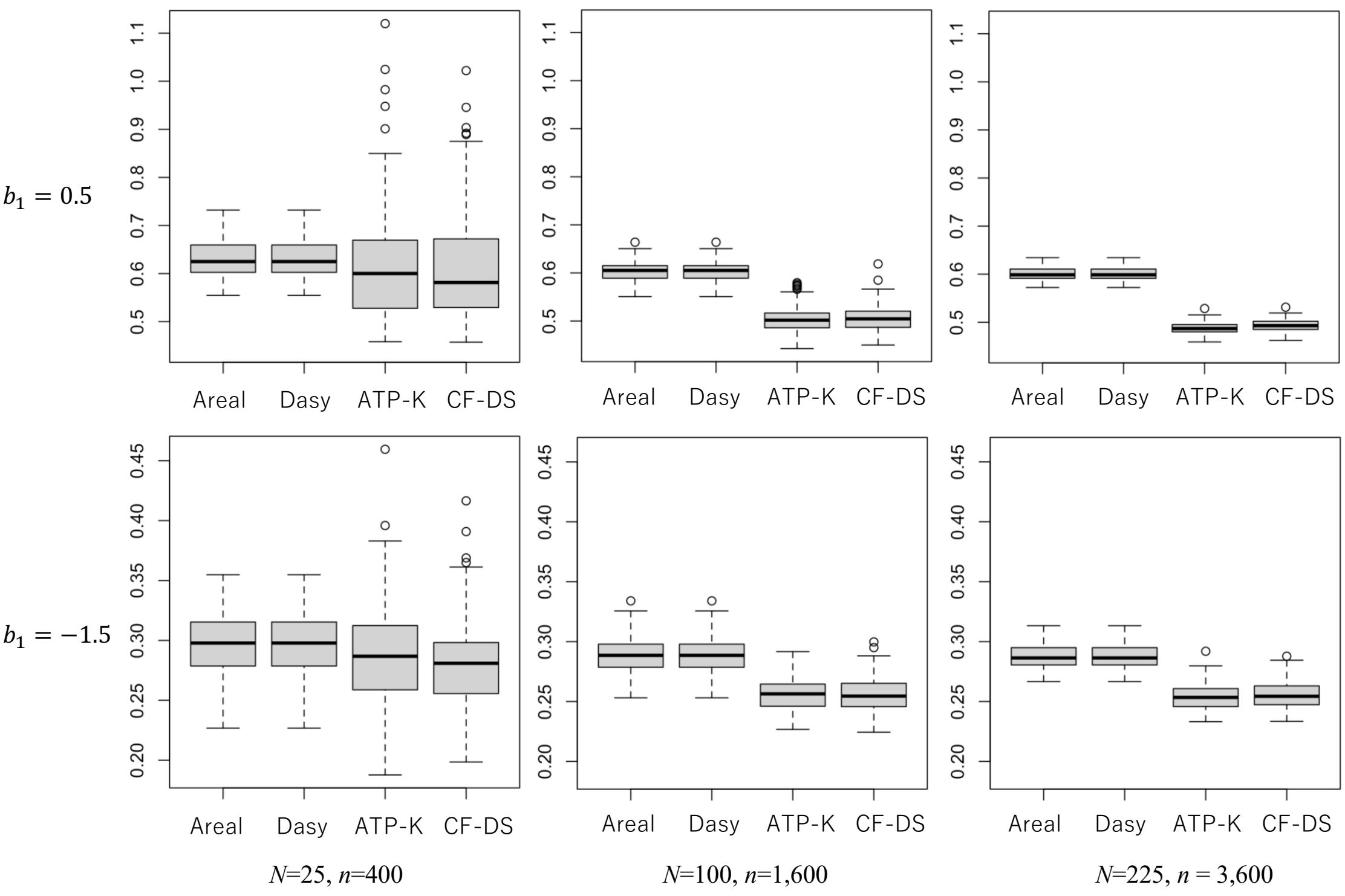

Figure A1: Boxplots of the RMSEs for intensive data.